\documentclass[11pt]{article}

\usepackage{amssymb}
\usepackage{bbold}
\usepackage{mathrsfs}
\usepackage{graphics}
\usepackage{color}

\newcommand{\be}{\begin{equation}}
\newcommand{\ee}{\end{equation}}
\newcommand{\bee}{\begin{eqnarray}}
\newcommand{\eee}{\end{eqnarray}}
\newcommand{\eqn}[1]{(\ref{eq:#1})}

\newcommand{\unitop}{\mathbb{1}}
\newcommand{\Reals}{\mathbb{R}}
\newcommand{\irrep}[1]{\mathbf{#1}}
\newcommand{\kahler}{\mathcal{K}}

\begin{document}


\begin{titlepage}

\title{
   \hfill{\normalsize hep-th/0409114\\}
   \vskip 2cm
   {\Large\bf $G_2$ Domain Walls in M-theory}\\[0.5cm]}
   \setcounter{footnote}{0}
\author{
{\normalsize\large Thomas House\footnote{email: thomash@sussex.ac.uk}
  \setcounter{footnote}{3}
  and Andr\'e Lukas\footnote{email: A.Lukas@sussex.ac.uk}} \\[0.5cm]
   {\it\normalsize Department of Physics and Astronomy, University of Sussex}\\
   {\normalsize Falmer, Brighton BN1 9QJ, UK} \\[0.2em] }
\date{}

\maketitle

\maketitle

\begin{abstract}\noindent
M-theory is considered in its low-energy limit on a $G_2$ manifold
with non-vanishing flux. Using the Killing spinor equations for linear
flux, an explicit set of first-order bosonic equations for
supersymmetric solutions is found. These solutions describe a warped
product of a domain wall in four-dimensional space-time and a deformed
$G_2$ manifold. It is shown how these domain walls arise from the
perspective of the associated four-dimensional $N=1$ effective
supergravity theories. We also discuss the inclusion of membrane and
M5-brane sources.
\end{abstract}

\thispagestyle{empty}

\end{titlepage}

\section{Introduction}

It is well-known that M-theory compactified on manifolds with $G_2$
holonomy leads to four-dimensional effective supergravities with $N=1$
supersymmetry~\cite{
Duff:1986hr,Papadopoulos:1995da,Duff:2002rw}. While such
compactifications on smooth $G_2$ spaces give rise to non-realistic
theories in four dimensions, simply consisting of Abelian vector
multiplets and uncharged chiral multiplets, it has been more recently
discovered that phenomenologically interesting theories can arise when
the $G_2$ space develops singularities~\cite{Atiyah:2001qf,Witten:2001uq,
Acharya:2001gy}. Accordingly, there has been considerable activity on
the subject~~\cite{Acharya:2000ps}--\cite{Lukas:2003rr} of M-theory
on $G_2$ spaces.

As most low-energy theories from string- or M-theory, the
four-dimensional effective theories from M-theory on $G_2$ spaces
contain a number of moduli fields, associated to the possible
deformations of the $G_2$ space. These moduli need to be fixed by a
low-energy potential to have a stable vacuum and any hope for
``realistic'' low-energy physics. It has been known for some
time~\cite{Dine:1985rz,Rohm:1985jv,Wen:1985qj} and has recently been studied in
more detail~\cite{Acharya:2002kv,Kachru:2003aw,Kachru:2003sx,Gukov:2003cy}, that
flux of anti-symmetric tensor fields is an effective tool to fix
moduli. Hence, $G_2$ flux compactifications constitute an interesting
framework for low-energy physics from M-theory.

\vspace{0.4cm}

It is these compactifications we wish to study in the present paper.
Generally, there are two somewhat complementary ways to approach flux
compactifications~\cite{Lukas:1998yy,Lukas:1998tt}.  Firstly, it can
be studied using the higher-dimensional theory by computing the
(supersymmetric) deformations of the $G_2$ background due to
non-vanishing flux. Typically one expects the flux to deform the $G_2$
space, introduce warping and modify the external four-dimensional
Minkowski space to a domain wall, as in the analogous case for
Calabi-Yau manifolds~\cite{Witten:1996mz,Lukas:1998yy,Lukas:1998tt}.
Examples of these domain wall solutions have been studied in
Refs.~\cite{Gibbons:2001ds,Stelle:2002dq}. A
systematic analysis of such flux backgrounds can be carried out by applying
the formalism of $G$ structures to M-theory compactifications~\cite{
Gauntlett:2002sc}--\cite{Lukas:2004ip}.

Alternatively, the problem can be approached from the viewpoint of the
four-dimensional effective supergravities which arise from a
flux-compactification on (un-deformed) $G_2$ spaces.  The general
structure of such theories, including a formula for the flux
superpotential, has been derived in Ref.~\cite{Beasley:2002db}. Due
to the presence of the non-trivial superpotential, the simplest
solution to these theories is not four-dimensional Minkowski space
but, rather, a domain wall.

The main goal of this paper is analyze $G_2$ flux compactifications
from both viewpoints and discuss the relation between them. On the
one hand, we will compute the supersymmetric deformation of the
11-dimensional $G_2$ background due to flux. This will be done to
linear order in flux, following the logic of the calculation in
Ref.~\cite{Witten:1996mz,Lukas:1997fg}. We will then consider the associated
four-dimensional $N=1$ supergravities and find their exact BPS domain
wall solutions. It is shown that these four-dimensional BPS domain
walls can be viewed as the zero-mode part of the full 11-dimensional
solution. We also demonstrate that the solutions can be supported by
either a membrane, located entirely in the external space, or an
M5-brane wrapping a three-cycle within the $G_2$ space.
 
\vspace{0.4cm}

We consider these results to be ``physically'' relevant in two ways.
Although, the $G_2$ domain walls do not respect four-dimensional
Poincar\'e invariance they may still provide a basis for
phenomenologically viable compactifications. This is because
non-perturbative effects which can be included in the four-dimensional
effective theory may yet produce a minimum of the 
potential~\cite{Dine:1985rz,Kachru:2003aw,Gorlich:2004qm,inprep}
and modify the domain wall to a four-dimensional maximally symmetric
space. More directly, our solutions represent the simplest way
in which a membrane (or a wrapped M5-brane) would appear in ``our''
four-dimensional universe if it indeed arises from $G_2$ compactification
of M-theory. In this sense, our results may provide the starting point
for an analysis of topological defects in an M-theory universe.

\vspace{0.4cm}

The plan of the paper is as follows. The next section sets up 
some basic equations and conventions and in Section \ref{sec:susyg2} we derive
the first-order differential equations describing the $G_2$
domain walls. In Section \ref{sec:explicit}, these equations are solved explicitly
in terms of a mode expansion on the $G_2$ space. The inclusion of
membrane and M5-brane sources is discussed in Section \ref{sec:brane}. In
Section \ref{sec:4d} we review the four-dimensional $N=1$ supergravities
from M-theory on $G_2$ spaces with flux and find their
domain wall solutions. These solutions are then compared to
their 11-dimensional counterparts. We conclude in Section \ref{sec:conc}.
Some technical information on gamma matrix conventions and
on $G_2$ group properties is collected in two appendices.

\section{Supergravity in 11 dimensions}

\label{sec:11dtheory}

In this section, we summarize our conventions for 11-dimensional
supergravity, following Ref.~\cite{Cremmer:1978km}, and set up some
notation. 
The bosonic part of the action is given by
\be \label{eq:11dsugra}
S_{11} = \frac{1}{2} \int_{X_{11}} d^{11} x \left[ \sqrt{-g} R
- \frac{1}{2} F\wedge *F - \frac{1}{6} A \wedge F \wedge F \right]\; ,
\ee
where $g$ is the 11-dimensional metric, $R$ is the associated Ricci
scalar and $A$ is a three-form field with field strength $F =
dA$. For simplicity, we have set the 11-dimensional Newton constant $\kappa_{11}$
to one. Throughout our calculations we take the expectation value of the
gravitino $\Psi$ to vanish. This means that we only need concern ourselves
with the bosonic equations of motion for this system, which are
\begin{eqnarray}
dF &=& 0 \label{eq:dF}\; ,\\
d^* F &=&- \frac{1}{2} F \wedge F \; ,\label{eq:dstarF}\\
R_{IJ} &=&\frac{1}{12}(F_{IKLM}F_{J}^{\ KLM}
- \frac{1}{12}g_{IJ} F\cdot F) \; . \label{eq:ricci}
\end{eqnarray}
Here $I,J\ldots$ are 11-dimensional space-time indices and
$R_{IJ}$ is the Ricci tensor. For $p$-forms $\zeta, \chi$, the notation
$\zeta \cdot \chi$ stands for the contraction of indices 
$\zeta_{i_1 \ldots i_p} \chi^{i_1 \ldots i_p}$.

The supersymmetric transformation of the gravitino is given by
\be
\delta_{\eta} \Psi_I =
D_I \eta + \frac{1}{288} F_{JKLM} \left(
{\Gamma_I}^{JKLM} - 8\delta_I^J \Gamma^{KLM} \right) \eta\; , \label{eq:gravver}
\ee
where $\eta$ is an 11-dimensional Majorana spinor and $D_I$ is the
spinor covariant derivative defined by 
\be
D_I = \partial_I + \frac{1}{4}
{\omega_I}^{\underline{I}\underline{J}}\Gamma_{\underline{I}\underline{J}} \; .
\ee
Underlining denotes tangent space indices, while multi-indexed symbols
$\Gamma^{I_1 \ldots I_p}$ denote anti-symmetrized products of
gamma-matrices, as usual. Our conventions for gamma-matrices are
explained in Appendix~\ref{app:spin}. It can be
shown~\cite{Gauntlett:2002fz,Gauntlett:2003wb} that a solution to the
Killing spinor equation $\delta_\eta\Psi_I=0$ which also satisfies the
form-field equation of motion~\eqn{dstarF} and the Bianchi
identity~\eqn{dF} provides a solution to the Einstein
equation~\eqn{ricci} as well.

\section{Finding supersymmetric $G_2$ domain wall solutions}

\label{sec:susyg2}

\subsection{General considerations}

In the absence of flux, the general M-theory backgrounds which lead to
four-dimensional $N=1$ supersymmetry consist of a direct product of a
$G_2$ manifold and four-dimensional Minkowski space. The main goal of
this paper is to understand how these backgrounds are modified in the
presence of flux. As is well-known~\cite{Beasley:2002db}, flux leads
to a non-vanishing moduli superpotential in the associated
four-dimensional effective theory. The ``simplest'' solution of this
theory is then a domain wall~\cite{Eto:2003bn} rather than
four-dimensional Minkowski space. We will return to this
four-dimensional viewpoint later. For the 11-dimensional Ansatz in the
presence of flux, this observation suggests we should accordingly
modify its four-dimensional Minkowski space part to a domain wall. As
we will see, also the metric on the $G_2$ space requires a correction
due to flux.

In practice, we will work with the Killing spinor equations, the
equation of motion~\eqn{dstarF} for $F$ and its Bianchi identity~\eqn{dF}.
To simplify the problem, the flux is regarded as an expansion parameter
and we will determine the flux-induced corrections to linear order.
The logic of the calculation is somewhat similar to one in 
Ref.~\cite{Witten:1996mz,Lukas:1997fg}
where flux corrections to Calabi-Yau backgrounds have been determined. 
The main result of this section will be a set of first order bosonic
differential equations for these linearized corrections.

\subsection{Covariantly constant spinors}

\label{Gcc}

As noted above, to obtain supersymmetric solutions we impose
that the variation of the gravitino \eqn{gravver} should vanish.
In the zero-flux regime, where we just consider the direct
product of Minkowski space with a $G_2$ manifold, $M_4 \times X_7$,
this amounts simply to imposing that there should be a covariantly constant
spinor, $\eta_0$, obeying
\be
D_I \eta_0 = \partial_I \eta_0 + \frac{1}{4}
{\omega_I}^{\underline{I}\underline{J}} 
\Gamma_{\underline{I}\underline{J}} 
\eta_0 = 0 \; .
\ee
In Appendix \ref{app:g2spin}, we explain why $G_2$ manifolds
will in general admit such a spinor.
When we introduce flux into the equation \eqn{gravver}, however, the
condition on the Killing spinor becomes more complicated. This
will lead us to perturb both the metric and spinor in order
to preserve supersymmetry.

\subsection{Metric Ansatz}

Following the earlier discussion, we shall consider solutions to M-theory
with line element corresponding to a warped product of an internal
seven-dimensional space and a domain wall in four dimensions, that is,
\be
ds^2 = e^{2\alpha} \eta_{\mu\nu} dx^{\mu} dx^{\nu}
+ e^{2\beta} dy^2 + g_{AB} dx^A dx^B \label{eq:metansatz} \; .
\ee
Here, we use indices $\mu , \nu \ldots = 0,1,2$, and $A,B \ldots = 4 \ldots 10$.
The three-dimensional part of the metric corresponds to the domain wall
worldvolume $X_3$ spanned by coordinates $x^\mu$ and $y$ is a coordinate transverse
to the wall. The seven-dimensional internal space $X_7$ with coordinates $x^A$
has a metric $g_{AB}$. Along with the warp factors $\alpha$ and $\beta$ it
generally depends on $x^A$ and $y$ but not on $x^\mu$. Therefore, we
have preserved three-dimensional Poincar\'e invariance on the
domain wall worldvolume, a general requirement which we will later
use to constrain the flux.

As we have explained, we would like to find a solution with metric
of the above form expanding to linear order in the flux. We should, hence, think
of the metric~\eqn{metansatz} as a linear perturbation of a direct product
of Minkowski space with a $G_2$ space with Ricci-flat metric. 
To this end, we expand to linear order in the warp factors $\alpha$ and
$\beta$ and write the internal seven-dimensional metric as
$g_{AB}=\Omega_{AB}+h_{AB}$, where $\Omega$ is a Ricci-flat
metric on a $G_2$ space and $h$ is the perturbation. The metric~\eqn{metansatz}
then takes the form
\be
ds^2 =  (1 + 2\alpha)\eta_{\mu\nu} dx^{\mu} dx^{\nu}
+ (1 + 2\beta)dy^2 + (\Omega_{AB} + h_{AB})dx^A dx^B 
\label{eq:linmetansatz}
\ee
Note that to zeroth order---that is setting $\alpha$, $\beta$ and $h$ to zero,
and in the absence of flux---this metric indeed provides a supersymmetric
solution to M-theory for the reasons given in Section \S \ref{Gcc}.
When we perturb the metric $g \mapsto g + h$, for linear $h$, 
we also perturb the spinor covariant 
derivative~\cite{Witten:1996mz} as
\be
D^I \mapsto D^I - \frac{1}{8} \left( 
\nabla_J h^I_{\ K} - \nabla_K h^I_{\ J}
\right) \Gamma^{JK} \; .
\ee
Hence, if we want Eq.~\eqn{gravver} to hold in the presence of flux,
we should think of the corrections $\alpha$, $\beta$ and $h$ 
as being ``sourced'' by flux. Our goal will be to determine their
explicit form as a function of the flux, such that the corrected solution
continues to preserve some supersymmetry.

\subsection{Conditions on the flux}

We will now write down the general form of the flux and the constraints
imposed on it by the $F$ equation of motion and the Bianchi identity.

Given we are asking for Poincar\'{e} invariance on the domain wall worldvolume
$X_3$, we are left with the following non-trivial components of $F$:
\be 
\begin{array}{rclrcl}
F_{ABCD} & = & G_{ABCD}\quad & F_{yABC} & = & J_{ABC} \\
F_{A\mu\nu\rho} & = & V_A \varepsilon_{\mu\nu\rho} &
F_{y\mu\nu\rho} & = & K \varepsilon_{\mu\nu\rho}\; .
\end{array} \label{eq:Fcomps}
\ee
Note that $G$, $J$, $V$, $K$ can be viewed as forms of various degree
on the internal space $X_7$.

Within the context of our expansion scheme, we consider flux as being first order.
At linear order, we can, therefore, neglect the $F\wedge F$ term in the
equation of motion \eqn{dstarF} and work with the zeroth order metric.
The $F$ equation of motion and the Bianchi identity then simply state that
\be \label{eq:fieldlin}
dF = d^* F = 0\; ,
\ee
where the Hodge star is with respect to the zeroth order metric. Inserting the various
component~\eqn{Fcomps} into Eq.~\eqn{fieldlin} we find from the Bianchi identity
\be
dG = dJ -  G' = dV = dK - V' = 0\; ,
\label{eq:Fbis}
\ee
and from the equation of motion
\be
d^*G - J' = d^*J = d^*V - K' = d^*K = 0\; .
\label{eq:Feoms}
\ee
Here a prime denotes differentiation with respect to $y$ and the
operators $d, d^*$ are now taken with respect to the internal space $X_7$
with Ricci-flat metric $\Omega$. To summarize, the most general flux is
described by a four-form $G$, a three-form $J$, a one-form $V$ and a scalar $K$
on the internal space $X_7$ which are subject to the equations~\eqn{Fbis}
and \eqn{Feoms}.

\subsection{Spinor Ansatz}

A somewhat delicate point in computations of the Killing spinor
equations is to find the most general Ansatz for the supersymmetry
spinor $\eta$. Sometimes, solutions to the Killing spinor equations
can be missed if, for example, a simple product Ansatz for $\eta$ is used.
We will, therefore, spend some time discussing this Ansatz for the
spinor and finding its most general structure. All relevant
conventions for spinors and gamma matrices in the various dimensions
involved are collected in Appendix \ref{app:spin}.

The first point to note is that $\eta$ must be
Majorana, that is
\be
\eta^{c} = \eta\; .
\ee
The conjugation is defined in Appendix \ref{app:spin}.
In general any such spinor can be written in terms of a Dirac
spinor $\psi$ like
\be
\eta = \psi + \psi^c \; .
\ee
If a pair of projectors $P_+, P_-$ can be found such that
\be
\begin{array}{ccc}
(P_{\pm})^2 = P_{\pm} & \left( P_{\pm} \psi \right)^c = P_{\mp} \psi^c
& P_+ + P_- = \unitop_{32} \; ,
\end{array}
\ee
then we may further write
\bee
\eta & = & \psi + \psi^c \nonumber \\
& = & P_+ (\psi + \psi^c) + P_- (\psi + \psi^c) \nonumber \\
& = & (P_+\psi + (P_-\psi)^c) + ((P_+\psi)^c + P_-\psi) \nonumber \\
& =: & \zeta + \zeta^c
\eee
where $\zeta = P_+ \zeta$.
Normally, $P_{\pm}$ would project onto positive and negative chiralities,
but there is no chirality operator in 11 dimensions so we do not yet have a
physical interpretation of the manipulation above. However, when we decompose
the spinor $\zeta$ as
\be
\zeta = \xi_+ \otimes \chi\; ,
\ee
where $\xi_+$ and $\chi$ are 4- and 7-dimensional spinors respectively,
we can define a sensible pair of projectors by
\be
P_{\pm} := \frac{1}{2} \left( 1 \pm \gamma \right) \otimes \unitop_8
\; . \ee 
This amounts to imposing that $\xi_+$ is a positive
chirality Weyl spinor, that is,  $\xi_+ = \gamma \xi_+$.
Its charge conjugate $\xi_- := \xi^c$ is then a
negative chirality spinor satisfying $\xi_- = - \gamma \xi_-$.
It is possible to show that, for our conventions, we have
\be
\begin{array}{cc}
\gamma^y \xi_+ = - (\xi_-)^* & \gamma^y \xi_- = (\xi_+)^*\; .
\end{array}
\ee
For an arbitrary complex number $z$ we have $z^* = e^{-2i\, \mathrm{arg}(z)} z$.
Of course, such a result will not in general hold for a multi-component complex
object like $\xi_+$. However, it turns out, after solving the Killing spinor
equations, there is no loss of generality in assuming that $\xi_+$ indeed
does satisfy such a relation. Consequently, we introduce a parameter $\theta$
such that
\be
\gamma^y \xi_{\pm} = e^{\pm i \theta} \xi_{\mp}\; .
\ee
Note that the internal part $\chi$ of the spinor remains unconstrained by the
projection, and at zeroth order in the flux will simply be the covariantly
constant spinor on the $G_2$ manifold, $\chi_0$. 

We now consider the perturbation of the spinor to linear order. For the
4-dimensional spinor we introduce a complex parameter $\epsilon$ such that
\be
\xi = (1+ \epsilon) \xi_+ \;
\ee
is the first order 4-dimensional spinor. Similarly to the argument about $\theta$ above,
this is not the most general perturbation of a multi-component complex
object, but there will be no loss of generality later if we take $\xi$ to be
of the above form.
We now use the results of Appendix \ref{app:g2spin}, to see that
the most general linear perturbation of the 7-dimensional spinor $\chi_0$ is given by
\be
\chi = (1 - v_0) \chi_0 + v^A \chi_A\; .
\ee
Here $v_0, v^A$ are complex variables parameterizing $\chi$.
In the full 11-dimensional picture, note that $\epsilon$ and $v_0$
are not really independent degrees of freedom since
we can absorb $\epsilon$ into a new parameter $v_{\epsilon} := \epsilon - v_0$.
We note here that this parameter encodes information about the variation of
$\theta$, since basic manipulation of the first order spinor gives
\be
\partial \theta = 2 \mathrm{Im} (\partial v_{\epsilon} ) \; . \label{eq:thetavar}
\ee
Because our conventions for the Dirac matrices allow us to express 11-dimensional 
charge conjugation as
\be
\zeta^c = \xi^c \otimes \chi^c\; ,
\ee
our final Ansatz for $\eta$ is then
\be
\eta = \xi_+ \otimes 
\left( (1 + v_{\epsilon}) \chi_0 + v^A \chi_A\right) + \xi_- \otimes 
\left( (1 + v_{\epsilon}^*) \chi_0^c + (v^A)^* \chi_A^c \right)\; .
\ee

\subsection{Bosonic equations}

Using the results above, together with the conventions of Appendix
\ref{app:spin} for the Dirac matrices and of Appendix \ref{app:g2}
for the action of these matrices on the $G_2$ spinor,
Eq.~\eqn{gravver} leads to a set of bosonic first-order equations. 
They constitute our main formal result and are given by
\bee
J \cdot \varphi & = & 12 K \cos \theta + \frac{1}{4} G \cdot \Phi \sin \theta
\label{eq:Kfirst} \\
12 V_A \cos \theta & = &
G_{ABCD} \varphi^{BCD} + J^{BCD} \Phi_{ABCD} \sin \theta \\
\partial_y \alpha & = & \frac{1}{144} G \cdot \Phi \cos \theta
- \frac{1}{3} K \sin \theta \label{eq:dyalpha} \\
\nabla_A \alpha & = & \frac{-1}{36} J^{BCD} \Phi_{ABCD} \cos \theta
- \frac{1}{3} V_A \sin \theta \label{eq:dAalpha} \\
\partial_y v_{\epsilon} & = & \frac{1}{288} \left( e^{- i \theta} G \cdot \Phi 
- 8iJ \cdot \varphi - 48i e^{- i \theta} K \right) \\
\nabla_A \beta & = & 2 \partial_y \left( \mathrm{Re} (v_A) \sin \theta
+ \mathrm{Im} (v_A) \cos \theta \right)
\nonumber \\ & &
+ \frac{1}{18} J^{BCD} \Phi_{ABCD} \cos \theta
+ \frac{1}{6} V_A \sin \theta \\
\partial_y \left( 
\begin{array}{r} \mathrm{Re} (v_A) \cos \theta
\\ + \mathrm{Im} (v_A) \sin \theta 
\end{array} \right) 
& = & \frac{-1}{72} \left(
G_{ABCD} \varphi^{BCD} 
- 2 J^{BCD} \Phi_{ABCD} \sin \theta + 6 V_A \cos \theta \right) \\
\nabla_A v_{\epsilon} & = & \frac{-1}{72}
\left( 2iG_{ABCD} \varphi^{BCD} 
+ e^{- i \theta } J^{BCD} \Phi_{ABCD} 
+ 12i e^{- i \theta } V_A \right) \\
\begin{array}{r}
4 \nabla_A v_{B} + i e^{- i \theta } \partial_y h_{AB}
\\ - \nabla_C h_{AD} \varphi^{CD}_{\ \ \ B}
\end{array}
 & = & 
\frac{1}{72} \left[ \begin{array}{l}
8 i G_{ACDE} \Phi^{CDE}_{\ \ \ \ \ B} 
+ \frac{i}{5} G^{CDEF} \big( 4\Phi_{ACDE} \Omega_{FB} \\
+ \Phi_{CDEF} \Omega_{AB} + 12 \varphi_{ACD} \varphi_{EFB}
+ 8 \varphi_{CDE} \varphi_{FAB} \big) \\
- 24 e^{- i \theta } J_{ACD} \varphi^{CD}_{\ \ \ B}
- 4 e^{- i \theta } J^{CDE} (
3 \varphi_{ACD} \Omega_{EB} - \\ \varphi_{CDE} \Omega_{AB} )
- 24 i e^{- i \theta } V^C \varphi_{ABC} + 24 K \Omega_{AB}
\end{array} \right]
\; .
 \label{eq:Klast}
\eee
Here $\varphi$ is the $G_2$ 3-form on the internal space, as defined in 
Appendix~\ref{app:g2lie} and $\Phi = *\varphi$ is its 4-form dual.

These equations link the parameters ($\alpha, \beta, h$), associated
with the metric, to the various flux components ($G,J,V,K$) and the
quantities ($v_{\epsilon}, v^A, \theta$) which parameterize the
Killing spinor. We note that the $y$-derivative of $\beta$ is
unconstrained by these equations, which is as we would expect from a
residual gauge degree of freedom after the choice of Ansatz. A
solution to these first-order partial differential equations preserves
two real supercharges ($N=1/2$ from a four-dimensional point of view)
and represents a warped product of a deformed $G_2$ space and a domain
wall in four-dimensional space-time.

\subsection{Ricci flatness}

It is a general result \cite{Gauntlett:2002fz,Gauntlett:2003wb} that the 
integrability of the Killing spinor
equation together with the field equations implies
that the Einstein equations hold. To linear order
in flux, this implies that our solutions should be
Ricci-flat. Let us confirm this by explicitly
computing the components of the Ricci tensor.
Using the Ansatz~\eqn{metansatz} we find
\bee
R_{\mu \nu} & = & (\partial_y^2 \alpha + \nabla_A^2\alpha )
 \eta_{\mu \nu} \nonumber \\
& = & \left( \frac{-1}{144} ( dJ - G') \cdot \Phi \cos \theta
- {1}{3} (K' - d^* V ) \sin \theta \right) \eta_{\mu \nu} \\
R_{yy} & = & 3\partial_y^2\alpha + \nabla_A^2\beta 
+ \frac{1}{2} \partial_y^2 h^A_{\ A} \nonumber \\ 
& = & \frac{1}{72} (dJ - G') \cdot \Phi \cos \theta +
\frac{1}{6} (K' - d^* V ) \sin \theta \\
R_{AB} & = & \nabla^A\nabla_B(3\alpha + \beta) + \frac{1}{2} \partial_y^2 h_{AB}
+ \nabla_A\nabla_{\left[ B \right. } h^C_{\ C\left. \right] }
+ \nabla^C\nabla_{\left[ C \right. } h_{B\left. \right] A}
- \frac{1}{2} h^C_{\ D} R^{D}_{\ \ ABC} \nonumber \\
& = & \frac{-1}{72} (dJ - G') \cdot \Phi \Omega_{AB} \cos \theta
+ \frac{1}{3} (K' - d^* V ) \Omega_{AB} \sin \theta
\nonumber \\ & & + \frac{1}{6} (dJ - G')_{(A}^{\ \ \ CDE} \Phi_{B)CDE} \label{eq:RAA} \\
R_{Ay} & = & \partial_y(3\nabla_A \alpha + \nabla_{\left[ A \right. } 
h^B_{\ B\left. \right] }) \nonumber \\
& = & - \frac{1}{72} (dG)_{BCDEF} \varphi_{A}^{\ \ BC} \varphi^{DEF}
- \frac{1}{12} (dV)_{BC} \varphi^{BC}_{\ \ \ A}\; .
\eee
With the first-order relations~\eqn{Kfirst}--\eqn{Klast}, together
with the conditions on the flux \eqn{Feoms} and \eqn{Fbis}, we find
that these components of the Ricci tensor indeed vanish.

\section{Explicit 11-dimensional solution}

\label{sec:explicit}

We now turn to the problem of integrating the bosonic equations
\eqn{Kfirst}--\eqn{Klast}. 
Since we are dealing with the case of a general
$G_2$ manifold, this solution will take the form of a sum over
basis sets of forms on the manifold. Although we have written
the bosonic equations above in the `raw' form in which
they are obtained, there is a certain amount of hidden gauge
symmetry that we would like to fix in our solution before
we write down such an expansion. In this section, we also consider
the zero-mode regime and its relation to compactification.

\subsection{Simplifying the spinor Ansatz}

Before we write down a solution to the 11-dimensional equations, we
reconsider the spinor ansatz. Our 7-dimensional spinor should be
invariant under ${\rm SO}(7)$ transformations of the tangent
space. Using the results of Appendix \ref{app:g2spin}
we can write such transformations as:
\be
\chi \mapsto e^{\theta^{AB} \Sigma_{AB}} \chi
= e^{\nu^A f_A + \mu^{AB} \rho_{AB}} \chi
\ee
where $\Sigma_{AB}$ are taken in the spinor representation of ${\rm SO}(7)$
and decompose into $f_A, \rho_{AB}$ under $G_2$ and
$\theta^{AB}$, $\nu^A$ and $\mu^{AB}$ are real parameters. To first
order, this  transformation reads
\be
\chi \mapsto (1 - v_0) \chi_0 + v^A \chi_A + \nu^A \chi_A\;
\ee
and means that we can `gauge away' Re$(v^A)$. The effects of a general
coordinate transformation on $\chi$ are similar but will
in general yield weaker conditions on $v^A$.

A further point to note is that, since we have a Killing spinor,
we are able to form bilinears in this spinor that will be
globally defined. In particular, the global vector $W^I = \bar{\eta} \Gamma^I \eta$,
formed in this way should itself be Killing~\cite{Gauntlett:2002fz}. 
At linear order, the transverse components of this vector are
\be
W^y = \cos \theta + 2( \mathrm{Im}(v_{\epsilon}) \sin \theta 
- \mathrm{Re}(v_{\epsilon}) \cos \theta )\; ,\qquad
 W^A = 4\, \mathrm{Im}(v^A)\; .
\ee
Since this vector must be Killing, we can then impose
\be
\nabla_{(A} v_{B)} = 0\; ,\qquad 
\partial_y  \mathrm{Im}(v_A) = 
\frac{1}{2} \nabla_A ( \mathrm{Re}(v_{\epsilon}) \cos \theta 
- \mathrm{Im}(v_{\epsilon}) \sin \theta )\; .
\ee
These relations allow us to eliminate $v^A$ from the
Killing spinor equation for $\beta$ which then takes the form
\be
\nabla_A \beta = 
\frac{1}{36} G_{ABCD} \varphi^{BCD} \sin \theta \cos \theta
+ \frac{3}{72} J^{BCD} \Phi_{ABCD} \cos \theta
+ \frac{1}{6} V_A \sin \theta\; .
\ee

\subsection{Simplifying the relations for metric perturbations}

We also make a gauge choice for $h_{AB}$, by putting
it in the standard `harmonic gauge' so that
\be
\nabla_B h^B_{\ A} = \frac{1}{2} \nabla_A \mathrm{tr} (h)\; . \label{eq:harmg}
\ee
Our result~\eqn{Kfirst}--\eqn{Klast} can be simplified considerably
by splitting into real and imaginary parts, projecting out
the irreducible $G_2$ representations associated with the two free
indices, using the simplifications of the spinor ansatz as above and making
the harmonic gauge choice for $h$. We are then able to derive the
following set of physically equivalent first order relations
\bee
J \cdot \varphi & = & 21 K \cos \theta + \frac{5}{8} G \cdot \Phi \sin \theta
\\
\partial_y \mathrm{tr} (h) & = & - \frac{5}{72} G \cdot \Phi \cos \theta
+ \frac{7}{3} K \sin \theta 
\\
\nabla_A \mathrm{tr} (h) & = & 4V_A \sin \theta - \frac{1}{3} J^{BCD} \Phi_{ABCD}
\\
\partial_y (P_{27} h)_{AB} - 
\nabla_C (P_{27} h)_{D(A} \varphi^{CD}_{\ \ \ B)} \sin \theta & = &
- \frac{1}{6} (P_{27} G)_{(A }^{\ \ \ CDE} \Phi_{B)CDE} \cos \theta
\\
\nabla_C (P_{27} h)_{D(A} \varphi^{CD}_{\ \ \ B)} \cos \theta & = &
- \frac{1}{6} (P_{27} G)_{(A }^{\ \ \ CDE} \Phi_{B)CDE} \sin \theta
\nonumber \\ & &
+ \frac{1}{2} (P_{27} J)_{(A}^{\ \ \ CD} \varphi_{B)CD}\; ,
\eee
where the projector $P_{27}$ projects out the ${\bf 27}$
representation in the $G_2$ decomposition of the various tensors, as
explained in Appendix \ref{app:project}.

\subsection{Zero-mode regime}

The field equations \eqn{Fbis} and \eqn{Feoms} imply that
\be
\begin{array}{cc}
\Delta_7 G = G'' & \Delta_7 J = J'' \\
\Delta_7 V = V'' & \Delta_7 K = K'' 
\end{array}
\ee
where $\Delta_7$ is the 7-dimensional Laplacian with respect to the zeroth order
metric $\Omega$. We call solutions
for which both sides of these equations are zero the
`zero modes' and those for which both are equal to a
non-zero constant the `massive modes'.

The physical reasoning behind this is that
operators like $\Delta_7$ will be associated with the inverse of the
radius of compactification of $X_7$. When this is reduced down to
small scales, this makes $\Delta_7$ produce extremely large
constant non-zero eigenvalues, which are effective masses in the 4-dimensional
theory. Since these masses will typically be at the Planck scale, they
can be ignored in constructing the 4-dimensional effective theory, and so the
zero-mode regime is of particular interest to us.

We firstly note that on $G_2$ manifolds there are no harmonic 1-forms,
and so the following terms in the flux vanish in the zero-mode regime:
\be
G_{ABCD} \varphi^{BCD} = J^{BCD} \Phi_{ABCD} = V_A = 0 \; .
\ee
This also constrains the spinor so that $v^A = 0$,
since otherwise the equation \eqn{Klast} would make $v^A$ a
harmonic 1-form.
Such constraints on 7-dimensional vectors mean that in the zero-mode
regime, using the first-order bosonic equations
\eqn{Kfirst}--\eqn{Klast} we have
\be
\nabla_A \alpha = \nabla_A \beta = 0 \; . \label{eq:daz}
\ee

A similar argument to that for the flux can be made for the
graviton $h_{AB}$, which from \eqn{RAA}, \eqn{harmg} and \eqn{daz} obeys
\be
\Box_7 h = - \partial_y^2 h\; .
\ee
where
\be
( \Box_7 h )_{AB} := 
\nabla_C \nabla^C h_{AB} + 2 R^C_{\ (AB)D} h^{D}_{\ C} \; .
\ee
In this case,
we also argue that $\Box_7$ will be associated with a Planck-scale effective
mass upon compactification and so can be ignored.

The arguments above allow us to write a `zero-mode' version of the first-order
11-dimensional bosonic equations \eqn{Kfirst}--\eqn{Klast}
\bee
\partial_y \alpha & = & \frac{1}{144} G \cdot \Phi \cos \theta
- \frac{1}{3} K \sin \theta \label{eq:zm1st}
\\
\partial_y \mathrm{tr} (h) & = & - \frac{5}{72} G \cdot \Phi \cos \theta
+ \frac{7}{3} K \sin \theta \label{eq:trhsimple}
\\
\partial_y (P_{27} h)_{AB} & = & - \frac{1}{6} 
(P_{27} G)_{(A }^{\ \ \ CDE} \Phi_{B)CDE} \cos \theta
\\
\partial_y \theta & = & - \frac{1}{48} G \cdot \Phi \sin \theta
- K \cos \theta
\\
P_{27} J & = & - P_{27} * G\; . \label{eq:zmlast}
\eee
We shall use these equations later to compare with the bosonic equations that
we derive from the 4-dimensional Killing spinor equations.

\subsection{Fourier expansion of the flux}

We shall now expand each component of the flux as a sum over forms on the
$G_2$ manifold $X_7$. 
At the zero-mode level, this expansion involves the harmonic
forms and is given by
\be
\begin{array}{rclrcl}
G_{0} & = & \sum_i G_i \psi^i \; ,\qquad & V_{0} & = & 0 \\
J_{0} & = & \sum_i J_i * \psi^i\; ,\qquad & K_0 & = & \mathrm{const.}
\end{array}
\ee
Here $G_i$, $J_i$ and $K_0$ are constants, and we have introduced
a set of harmonic 4-forms on $X_7$, $\{ \psi^i \}_{i = 1}^{b_3(X_7)}$ 
where $b_3(X_7)$ is the 3rd Betti number of $X_7$.
Notationally, we will sometimes adopt implicit summation over
$i, j$ type indices but leave them in for clarity at present.

In the massive regime, the expansion is slightly more complicated, since
we must introduce a further set of massive 4-forms on $X_7$, $\{ \Psi^n \}$ 
satisfying
\be
\Delta_7 \Psi^n = (m_n)^2 \Psi^n\; . 
\ee
We can then use the Hodge star to construct a set of 3-forms,
$\{ * \Psi^n \}$, with
\be
\Delta_7 * \Psi^n = (m_n)^2 * \Psi^n\; . 
\ee
Then the massive modes of $G,J$ can be expanded in terms of these forms,
leading to
\be
\begin{array}{cc}
G_{\mathrm{massive}} = \sum_n G_n(y) \Psi^n 
& J_{\mathrm{massive}} = \sum_n J_n(y) * \Psi^n\; ,
\end{array}
\ee
with $y$-dependent expansion coefficients $G_n$ and $J_n$.
The equations of motion for the flux then imply
\be
\begin{array}{rclcrcl}
G_n'' & = & (m_n)^2 G_n 
& \Rightarrow &
G_n(y) & = & G_n^+ e^{m_n y} + G_n^- e^{-m_n y}
\\
J_n'' & = & (m_n)^2 J_n 
& \Rightarrow &
J_n(y) & = & J_n^+ e^{m_n y} + J_n^- e^{-m_n y}\; ,
\end{array}
\ee
for constant $G_n^+,G_n^-,J_n^+, J_n^-$.
The massive expansion of $K$ and $V$ can be done in a similar way.
We can write both in terms of a set of functions $\left\{ f^p \right\} $ obeying
$ \Delta_7 f^p = (M_p)^2 f^p $ so that
\bee
V_{\mathrm{massive}} & = & 
\sum_p
\frac{1}{M_p} \left( V_p^+ e^{M_p y} + V_p^- e^{-M_p y} 
\right) df^p \nonumber \\
K_{\mathrm{massive}} & = & 
\sum_p
\left( K_p^+ e^{M_p y} + K_p^- e^{-M_p y} \right) f^p
\eee
for constants $V_p^+$, $V_p^-$, $K_p^+$ and $K_p^-$.
We have introduced a
factor of $M_p$ in the first of these relations to compensate for the
mass associated with the exterior derivative $d$.
To see why this is the correct expansion for our solution
consider the linear equation for $\nabla_A \alpha$, \eqn{dAalpha},
\bee
\nabla_A \alpha & = &
\frac{-1}{36} J^{BCD} \Phi_{ABCD} \cos \theta
- \frac{1}{3} V_A \sin \theta  = \nonumber \\
(d \alpha)_A & = & \frac{-1}{36} \sum_n
(J_n^+ e^{m_n y} + J_n^- e^{-m_n y})
(* \Psi^n)^{BCD} \Phi_{ABCD} \cos \theta \nonumber \\ & &
- \frac{1}{3} \sum_p
\left( V_p^+ e^{M_p y} + V_p^- e^{-M_p y} \right)
(df^p)_A \sin \theta
\eee
which, from the nilpotency of $d$ implies that the right hand side of this
equation is a closed 1-form and thus can be written uniquely as the sum of an
exact 0-form and a harmonic 1-form. As there are no harmonic 1-forms on
$X_7$ due to $G_2$ holonomy, each term on the right hand side must be exact.
This justifies our expansion of $V$ in terms of the $\left\{ d f^p \right\} $,
and also allows us to define a further set of functions by
\be
( d \Psi_{(0)}^n )_A := m_n (* \Psi^n)^{BCD} \Phi_{ABCD}\; .
\ee

\subsection{Integration of the bosonic equations}

After expanding as above, the direct integration of the bosonic equations 
\eqn{Kfirst}--\eqn{Klast} is relatively straightforward, particularly
as we can take $\theta$ as a constant to linear order. Performing
this integration then leads
the following complete solution for the metric components
\bee
\alpha(y, x^A ) & = & \left( \frac{-1}{3} K_0 y
- \frac{1}{3} \sum_p \frac{1}{m_p} \left( 
(K_p^+ + V_p^+) e^{m_p y} - (K_p^- - V_p^-) e^{- m_p y}
\right) f^p \right) \sin \theta  = \nonumber \\
& & + \frac{1}{144} \left( \sum_i G_i \psi^i y + \sum_n
\frac{1}{m_n}( G_n^+ e^{m_n y} - G_n^- e^{-m_n y} ) \Psi^n \right)
\cdot \Phi \nonumber \cos \theta \\ & &
- \frac{1}{36} \sum_n \frac{1}{m_n} \left (J_n^+ e^{m_n y} + J_n^- e^{-m_n y}\right) 
\Psi_{(0)}^n \cos \theta + \alpha_0
\\
\beta(y, x^A) & = & \frac{1}{36} \left( \sum_i G_i \psi^i y + \sum_n
\frac{1}{m_n}( G_n^+ e^{m_n y} - G_n^- e^{-m_n y} ) \Psi^n \right)
\,_{ABCD} \varphi^{BCD}
\sin \theta \cos \theta \nonumber \\ & & 
+ \frac{1}{6} \sum_p \frac{1}{m_p} \left( 
V_p^+ e^{m_p y} + V_p^- e^{- m_p y}
\right) f^p \sin \theta \nonumber \\ & &
+ \frac{3}{72} \sum_n \frac{1}{m_n} 
(J_n^+ e^{m_n y} + J_n^- e^{-m_n y}) \Psi_{(0)}^n \cos \theta 
+ \beta_0 \\
\mathrm{tr} (h) (y, x^A) & = & \left( \frac{7}{3} K_0 y +
\sum_p \frac{1}{m_p} \left( 
\left (\frac{7}{3}K_p^+ + 4V_p^+\right) e^{m_p y} 
- \left(\frac{7}{3}K_p^- - 4V_p^-\right) e^{- m_p y}
\right) f^p \right) \sin \theta \nonumber \\
& & - \frac{5}{72} \left( \sum_i G_i \psi^i y + 
\sum_n \frac{1}{m_n}( G_n^+ e^{m_n y} - G_n^- e^{-m_n y} ) \Psi^n \right)
\cdot \Phi \cos \theta \nonumber \\  & & 
- \frac{1}{3} \sum_n \frac{1}{m_n} (J_n^+ e^{m_n y} 
+ J_n^- e^{-m_n y}) \Psi_{(0)}^n
\cos \theta + h_0\; .
\eee
Here, $\alpha_0, \beta_0, h_0$ are constants of integration and we
have gauged away the $y$-dependence of $\beta$. Note that the first
term, proportional to $y$ on each right-hand side represents the
zero-mode contributions while all other terms correspond to heavy
modes.  The above expressions make it clear that the massive modes are
proportional to the inverse effective mass, and so upon
compactification we expect them to be negligible. This means, we
should consider only the zero mode regime when we come to the
4-dimensional effective theory later.

For the traceless part of the metric, the integration is slightly
less straightforward, but proceeds along similar lines so that
at zero mode
\be
(P_{27} h_{\mathrm{zero-mode}})_{AB} = -\frac{1}{6} \sum_i
G_i y \cos \theta (P_{27} \psi^i)_{(A }^{\ \ \ CDE}  \Phi_{B)CDE} \; .
\ee
The massive modes can be obtained by solving the equations
\bee
\lefteqn{
\partial_y (P_{27} h_{\mathrm{massive}} )_{AB} - 
\nabla_C (P_{27} h_{\mathrm{massive}})_{D(A} \varphi^{CD}_{\ \ \ B)} \sin \theta 
= } \nonumber \\ & &
-\frac{1}{6} \sum_n
\left( G_n^+ e^{m_n y} + G_n^- e^{-m_n y} 
\right) ( P_{27} \Psi^n )_{(A }^{\ \ \ CDE}  \Phi_{B)CDE}  \cos \theta 
\nonumber \\
\lefteqn{
\nabla_C (P_{27} h_{\mathrm{massive}} )_{D(A} \varphi^{CD}_{\ \ \ B)} \cos \theta
= } \nonumber \\ & & 
-\frac{1}{6} \sum_n
\left( ( G_n^+\sin \theta  + J_n^+) e^{m_n y} + 
(G_n^- \sin \theta  + J_n^-) e^{-m_n y} 
\right) ( P_{27} \Psi^n )_{(A }^{\ \ \ CDE}  \Phi_{B)CDE} \; , 
\eee
which can be done explicitly. 

\subsection{Curvature singularities}

Analogously to the calculation in Ref.\ \cite{Witten:1996mz}, we now
look for the curvature singularities that correspond to vanishing volume of the
compact space. This quantity is given by
\be
V := \mathrm{Vol} (X_7) = \int_{X_7} \sqrt{g_7}\, d^7 x\; .  \label{eq:defVol}
\ee
We can then differentiate this to linear order using the relation \eqn{trhsimple}
so that
\bee
\partial_y V & = & \int_{X_7} \frac{1}{2} \Omega^{AB} 
\partial_y h_{AB} \sqrt{\Omega} d^7 x \nonumber \\
& = & \int_{X_7} \left( - \frac{5}{144} G \cdot \Phi \cos \theta
+\frac{7}{6}  K \sin \theta \right) \sqrt{\Omega} d^7 x \nonumber \\
& = & \int_{X_7} \left( - \frac{5}{6} \varphi \wedge G \cos \theta
+\frac{7}{6} * K \sin \theta \right) \label{eq:dyVol} \; .
\eee
We can then use the equations of motion for the flux \eqn{Fbis} 
and \eqn{Feoms} to show that
\bee
\partial_y^2 V & = & \int_{X_7} d \left( - \frac{5}{6}
(\varphi \wedge J) \cos \theta + \frac{7}{6} (*V) \sin \theta
\right) \nonumber \\
& = & 0 \label{eq:dydyVol} \; ,
\eee
since $X_7$ has no boundary. The volume must thus depend linearly on
the coordinate $y$, and so it must be zero for some value of $y$,
which will correspond to a curvature singularity of the internal
space.  Of course, as the volume of the compact space becomes small,
we are no longer entitled to use simply 11-dimensional supergravity as
our theory. We might also reasonably expect that although the linear
terms in flux in \eqn{dydyVol} vanish, the higher-order contributions
may not.

\section{Inclusion of brane sources}

\label{sec:brane}

In general, we expect $(p+1)$-form fields to be sourced by
an extended charged $p$-brane. 
In M-theory, there are two sensible choices for $p$, given that the
only form field present is the three-form field $A$. These are the
`fundamental' membrane (M2-brane) and the `magnetic' five-brane (M5-brane).

We shall consider each of these in turn, as both may well support
the kind of domain wall solution that we are considering. In the
case of the M2-brane, this could happen by simply sitting in the
external space, whereas the M5-brane would have to wrap a three-cycle
in the compact space in an appropriate way.

Our approach shall be to solve the brane equations of motion for
each system, and then try to match these solutions to
appropriate specializations of the bulk solution that we have so
far been considering. In doing this, we will find that
the inclusion of a brane source fixes the value of $\theta$
in such a way that our bulk solution can either support the
M2-brane or the M5-brane but not both.

We further find that in each case the brane splits the $y$-direction
into two regions, each with different values of the flux and
with a `jump' in certain components of the flux across the
brane proportional to the brane tension.

\subsection{General brane action in M-theory}

In this section, we will quote some general results about classical
membranes in 11-dimensional supergravity, following Ref.~\cite{Duff:1994an}. 
The easiest to consider is the fundamental membrane, which
couples to the 3-form field $A$ by the action
\be
S_{2} = T_{2} \int_{W_{3}} d^{3} \mathcal{X} \Bigg( -{1\over 2}
\sqrt{-\gamma}
 \gamma^{ij}
\partial_i X^I \partial_j X^J g_{IJ} -
\frac{1}{2}
\sqrt{-\gamma}
 - \frac{1}{4!} \varepsilon^{i j k} \partial_{i} X^{I}
\partial_{j} X^{J} \partial_{k} X^{K} A_{IJK} \Bigg) \; ,
\ee
where $X^I$ are the brane coordinates, $i,j, \ldots = 0, \ldots , 2$ are 
brane worldvolume indices and $W_{3}$ is the worldvolume of the brane parameterized by
the coordinates $\mathcal{X}^i$.
This membrane couples to our bulk
SUGRA action $S_{11}$ in Eq.~\eqn{11dsugra} to produce the total action
\be
S_{\mathrm{Total}} = S_{11} + \int d^{11} x\, S_2 \delta (X - x) \; .
\ee
This modifies the previous equations of motion to give
\bee
dF & = & 0 \\
d^* F & = & -2 \mathcal{J}_2
\\
R_{IJ} & = & \frac{1}{12}\left( F_{IKLM}F_{J}^{\ KLM}
- \frac{1}{12}g_{IJ} F\cdot F\right) + \sqrt{-g}
\left( T_{IJ} - \frac{1}{9} g_{IJ} T \right) \; .
\eee
The current and stress-energy associated with the brane are
\bee
\mathcal{J}_2^{I J K} & = & T_2 \int d^{3} \mathcal{X}
\varepsilon^{i j k}
\partial_{i} X^{I} \partial_{j} X^{J} \ldots \partial_{k}
X^{K}{\delta^{11} (x-X)\over \sqrt{-g}}
\\
T^{IJ} & = & -T_2 \int d^{3} \mathcal{X}
\sqrt{-\gamma}
 \gamma^{ij} \partial_i
X^I \partial_j X^J {\delta^{11} (x-X)\over \sqrt{-g}} \; ,
\eee
and the membrane worldvolume equations of motion are
\bee
\partial_i \left( \sqrt{-\gamma} \gamma^{ij}
\partial_j X^J
g_{IJ}\right) & = & {1\over 2} \sqrt{-\gamma} \gamma^{ij}
\partial_i X^J
\partial_j X^K\partial_I g_{JK} 
+ \frac{1}{4!} \varepsilon^{i j k} \partial_{i} X^{J}
\partial_{j} X^{K} \partial_{k} X^{L}
F_{I J K L} 
\\
\gamma_{ij} & = & \partial_i X^I \partial_j X^J g_{IJ} \; .
\eee

The coupling of the five-brane to the bulk action is slightly more
subtle, and to be done properly requires the approach of 
\cite{Bandos:1997gd}. As we are only interested in the effect of
the five-brane on the equations of motion rather than the duality-symmetric
formulation of the low-energy M-theory action, we merely state here 
the five-brane equations of motion when we set the worldvolume
two-form to zero. Analogously to the magnetic
monopole in electrodynamics, the five-brane sources to the dual of
the flux, $*F$, in the same way that the membrane couples to $F$.
We then have bulk equations of motion
\bee
d^* F & = & 0 \\
dF & = & 2 * \mathcal{J}_5
\\
R_{IJ} & = & \frac{1}{12}\left( F_{IKLM}F_{J}^{\ KLM}
- \frac{1}{12}g_{IJ} F\cdot F\right) + \sqrt{-g}
\left( T_{IJ} - \frac{1}{9} g_{IJ} T \right) \; .
\eee
In this case the current and stress-energy are
\bee
\mathcal{J}_5^{I_1 \ldots I_6 } & = & T_5 \int d^{6} \mathcal{X}
\varepsilon^{i_1 \ldots i_6}
\partial_{i_1} X^{I_6} \ldots \partial_{i_6} X^{I_6}
{\delta^{11} (x-X)\over \sqrt{-g}}
\\
T^{IJ} & = & -T_5 \int d^{6} \mathcal{X}
\sqrt{-\gamma}
 \gamma^{ij} \partial_i
X^I \partial_j X^J {\delta^{11} (x-X)\over \sqrt{-g}} \; ,
\eee
with worldvolume equations
\bee
\partial_i \left( \sqrt{-\gamma} \gamma^{ij}
\partial_j X^J
g_{IJ}\right) & = & {1\over 2} \sqrt{-\gamma} \gamma^{ij}
\partial_i X^J
\partial_j X^K \partial_I g_{JK} 
+ \frac{1}{7!} \varepsilon^{i_1 \ldots i_6} \partial_{i_1} X^{I_6}
\ldots \partial_{i_6} X^{I_6}
(* F)_{I_1 \ldots I_6} 
\\
\gamma_{ij} & = & \partial_i X^I \partial_j X^J g_{IJ} \; .
\eee

We shall now go on to apply these results for the M2-brane and M5-brane in
the context of our existing bulk solution, noting that
from the brane equations the flux is naturally of the order
of the brane tension, which we must bear in mind when we truncate our
results to linear order.

We will then go on to solve the membrane equations of motion at linear order,
but only considering the zero-mode part, which will involve `smoothing
over' the 7-dimensional part of the delta functions in the action, essentially
for reasons of simplicity. There is no clear reason why a fuller treatment
of the massive modes should not yield fundamentally the same
conclusions as below.

\subsection{Fundamental membrane}

It is natural to consider the membrane as supporting a domain wall
solution by placing it in the external space with its transverse
coordinate along the special direction, previously called $y$,
of the solution. We implement this with the following `static gauge'
choice for the membrane coordinates
\be
X^{\mu} = \xi^{\mu}\; ,\qquad X^y = \mathrm{const.}\; ,\qquad X^A = \mathrm{const.}
\ee
If we then consider the linearized 11-dimensional equations of motion, taking only the zero
mode of the internal space part, we find the following modifications
\bee
\partial_y^2 \alpha  & = & \frac{-2}{3} T_2 \delta(y) \label{eq:memb1} \\
\partial_y^2 \mathrm{tr} (h) & = & \frac{14}{3} T_2 \delta(y) \label{eq:memb2} \\
K' & = & 2 T_2 \delta(y) \label{eq:memb3} \; .
\eee
The membrane worldvolume equations are
\be
\begin{array}{cc}
\partial_y \alpha = -\frac{1}{3}K & \nabla_A \alpha = -\frac{1}{3} V_A \; ,
\end{array}
\ee
and imposing worldvolume supersymmetry ($\kappa$-symmetry) gives the
following condition:
\be
\widetilde{P_-} \eta = 0
\ee
where $\widetilde{P_{\pm}} := \frac{1}{2} (1 \pm i \gamma^y \gamma) \otimes \unitop$
can be interpreted as projecting out
different components of 8-D chirality. This condition turns out to be equivalent
to taking $\sin \theta = 1$ in the Killing Spinor ansatz.

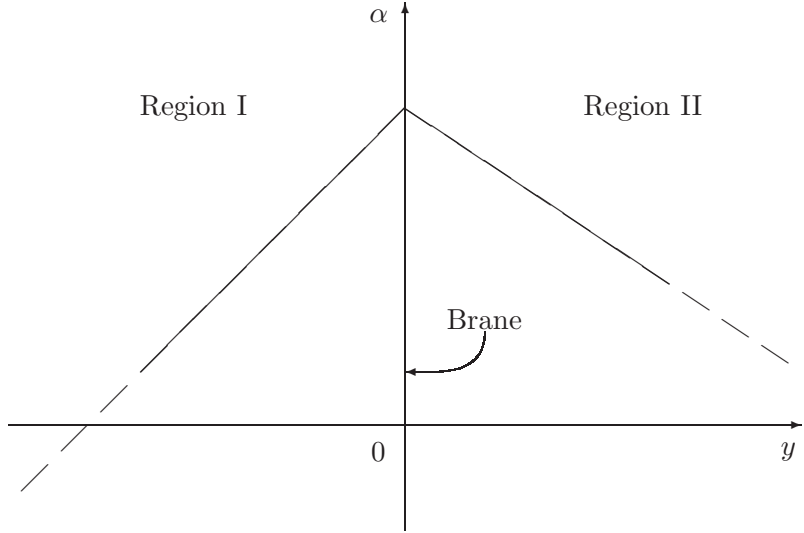
\begin{figure}

\begin{center}

\begin{picture}(300,200)(0,0)

\put(0,40){\vector(1,0){300}}
\put(150,0){\vector(0,1){200}}

\put(140,195){\makebox(0,0){$\alpha$}}
\put(295,30){\makebox(0,0){$y$}}
\put(140,30){\makebox(0,0){$0$}}

\put(150,160){\line(-1,-1){100}}
\multiput(150,160)(-15, -15){10}{\line(-1,-1){10}}
\put(150,160){\line(3,-2){100}}
\multiput(150,160)(15, -10){10}{\line(3,-2){10}}

\put(180,80){\makebox(0,0){Brane}}
\qbezier(180,75)(180,60)(160,60)
\put(160,60){\vector(-1,0){10}}

\put(70,160){\makebox(0,0){Region I}}
\put(240,160){\makebox(0,0){Region II}}

\end{picture}

\end{center}

\caption{Physical Interpretation of the Membrane Solution \label{fig:brane}}
\end{figure}

The physical interpretation of this solution is shown in Figure \ref{fig:brane}.
Having smoothed over the compact space, we can consider the remaining 4-dimensional
space to be split into two regions by the membrane, each containing
different (constant) values for the flux.

Suppose we write the flux in Region I as $K_{\mathrm{I}}$ 
and in Region II as $K_{\mathrm{II}}$, then we can integrate the equations
\eqn{memb1}-\eqn{memb3} by writing
\be
K(y) = 2T_2 (K_{\mathrm{II}} - K_{\mathrm{I}} ) \Theta(y) + K_{\mathrm{I}}
\ee
\be
\begin{array}{cc}
\partial_y \alpha = -\frac{1}{3}K &
\partial_y \mathrm{tr} (h) = \frac{7}{3}K
\end{array}
\ee
where $\Theta$ is the step function defined by
\be
\Theta(x) := \left\{ \begin{array}{c} 
0 \mathrm{\ for\ } x < 0  \\
\frac{1}{2} \mathrm{\ for\ } x = 0 \\
1 \mathrm{\ for\ } x > 0 \; .
\end{array} \right.
\ee
This solution is consistent with taking the $\sin \theta = 1$ specialization
of the zero-mode bosonic equations \eqn{Kfirst}--\eqn{Klast} above in each of the bulk
regions. The `jump' in flux from one region to another is proportional
to the membrane tension. It is worth noting that membranes could be
stacked to make this jump proportional to the number of membranes
times the tension.

\subsection{Magnetic five-brane}

The situation for the M5-brane is slightly more complicated than for the M2-brane,
given that it has three more worldvolume dimensions than the domain wall.
For this reason, three of the worldvolume dimensions should be wrapped
on some three-cycle $\Sigma_3$ with in the $G_2$ space $X_7$.
For the solution that arises from this configuration to be supersymmetric,
this cycle must be \emph{calibrated} by the $G_2$ three-form $\varphi$,
which means that
\be
\mathrm{Vol} (\Sigma_3) = \int_{\Sigma_3} \varphi \; .
\ee
The appropriate choice for the brane coordinates is then
\be
X^{\mu} = \xi^{\mu}\; ,\qquad X^y = \mathrm{const.}\; ,\qquad
X^a = \sigma^a\; ,\qquad X^{\tilde{A}} = \mathrm{const.} \; ,
\ee
where we let $\sigma^a$ be the coordinates of $\Sigma_3$ and use
the indices $\tilde{A}$ to denote directions perpendicular to $\Sigma_3$.
Considering the equations of motion for this system in the zero mode
regime then gives
\bee
\partial_y^2 \alpha  & = & \frac{-1}{3} T_5 \delta(y) \label{eq:m51} \\
\partial_y^2 \mathrm{tr} (h) & = & \frac{10}{3} T_5 \delta(y) \label{eq:m52} \\
G' & = & \frac{2}{7} T_5 \delta(y) \Phi \label{eq:m53} \; ,
\eee
with worldvolume equations
\be
6 \partial_y \alpha + \frac{3}{7} \partial_y \mathrm{tr} (h) =
\frac{-1}{84} G \cdot \Phi
\ee
Imposing $\kappa$-symmetry gives the following condition:
\be
P_-^y \eta = 0
\ee
where $P^y_{\pm} := \frac{1}{2} (1 \pm \gamma^y ) \otimes \unitop $ 
can be interpreted as projecting out
different components of $y$-chirality. This condition turns out to be equivalent
to taking $\cos \theta = 1$ in the Killing Spinor ansatz.

Our 4-dimensional picture then looks very similar to that of the membrane,
with two separated regions containing different values for the
flux, such that
\be
G(y) = 2T_5 (G_{\mathrm{II}} - G_{\mathrm{I}} ) \Theta(y) + G_{\mathrm{I}} \; .
\ee
Note that only the singlet part of $G$ is shifted by the brane
since, from Eq.~\eqn{m53}, the difference  $G_{\mathrm{II}} - G_{\mathrm{I}}$
is proportional to the $G_2$ invariant four-form $\Phi$.
As a result, we need not consider the traceless part of $h$ and can thus
simply integrate \eqn{m51} and \eqn{m52} to give
\be
\partial_y \alpha = \frac{-1}{144} G \cdot \Phi\; ,\qquad
\partial_y \mathrm{tr} (h) = \frac{5}{72} G \cdot \Phi \; .
\ee
Similarly to the M2-brane case, this is consistent with the $\cos \theta = 1$
specialization of the bosonic equations \eqn{Kfirst}--\eqn{Klast}.
We also have the partition of the external space into two
separate regions each with different constant values for the
flux, with the jump in flux between these regions proportional
to the brane tension. In contrast to the M2-brane, however, the
relevant component of flux is the singlet of $G$ rather than
the $K$. It will also be possible to stack branes so that
the jump in flux is proportional to the number of stacked
branes.

Note that although both the membrane and five-brane very naturally
couple to our bulk solution, at the order we are considering, the
existence of supersymmetric configurations with two real supercharges
containing both types of brane is ruled out.

\section{Four-dimensional effective theory}

\label{sec:4d}

When 11-dimensional SUGRA is compactified on a $G_2$ manifold, the
effective field theory is given by 4-dimensional $N=1$ SUGRA. In this
section, we outline the action for this theory, together with how the
quantities in that action are related to the 11-dimensional
quantities.  We then present the conditions for an N=1 supersymmetric
domain wall solution in 4 dimensions, and integrate these
equations. Finally, we uplift the 4-dimensional equations to 11
dimensions, and check that the result indeed matches our earlier one,
obtained directly from the 11-dimensional theory.

\subsection{For-dimensional $N=1$ supergravity}

The relevant bosonic terms in 4-dimensional $N=1$ supergravity action are
\be
S_4 = - \frac{1}{2} \int \left( \sqrt{-g} R
+ 2 \mathcal{K}_{i \bar{\jmath}} \partial_m T^i
\partial^m \overline{T^{\jmath}} + 2 U \right)
\ee
where $m = 0 \ldots 3$. As we have done with its 11-dimensional counterpart,
we have set the four-dimensional Newton constant $\kappa_4$ to one.
The fields $T^i$ are scalar components of chiral superfields, and
$\mathcal{K}_{i \bar{\jmath}}$ is the Kahler metric, given by
\be
\kahler_{i\bar{\jmath}} =
\frac{\partial^2 \kahler}{\partial T^i \partial \overline{T^{\jmath}}}
\ee
in terms of the Kahler potential $\kahler$.
Field indices $i,j,\dots $ are lowered and raised by $\kahler_{i\bar{\jmath}}$
and its inverse $\kahler^{i\bar{\jmath}}$.
The potential $U$ is given in terms of the
superpotential $W$ by
\be
U = e^{\kahler} \left( \kahler^{i\bar{\jmath}} D_i W 
D_{\bar{\jmath}} \overline{W} -3|W|^2 \right)
\ee
where the Kahler covariant derivative $D_i$ is defined by
\be
D_i := \frac{\partial}{\partial T^i} + \kahler_i \; .
\ee

\subsection{4-dimensional SUGRA from M-Theory on a $G_2$ space}

The relationship between the theory above and 11-dimensional SUGRA 
on a compact $G_2$ manifold is covered
in Refs.~\cite{Papadopoulos:1995da} and \cite{Beasley:2002db}.
Throughout this section we use just the zero-mode part of form field
strength $G$ which is written in terms of a set of harmonic 4-forms
$\{ \psi^i \}$ as
\be
G = G_i \psi^i
\ee
where we now use implicit summation rather than the
explicit notation of Section \ref{sec:explicit}.
Our first step is to expand both the 3-form field $A$
and the $G_2$ 3-form $\varphi$
in terms of a dual set of harmonic
three-forms $\{ \pi_i \}_{i = 1}^{b_3(X_7)}$ obeying
\be
\int_{X_7} \pi_i \wedge \psi^j = \delta_i^j
\ee
so that
\bee
\varphi & = & \varphi^i \pi_i 
\\
A & = & A^i \pi_i + G_i \tau^i
\eee
where $d \tau^i = \psi^i$, so that $F = dA$ still holds.
Detailed consideration of the compactification of the 11-dimensional
theory shows that
the $\varphi^i$ are the metric moduli of the $G_2$ manifold, $X_7$,
and the moduli $A^i$ appear as axions in the 4-dimensional theory.
This means that we can write the superfields as
\be
T^i = \varphi^i + i A^i \; .
\ee
In our conventions, the superpotential is then
\bee
W & = & \frac{7^{3/2}}{4} \int_{X_7} 
\left(\varphi + \frac{i}{2} A\right) \wedge G \label{eq:W1} \\
& = & \frac{7^{3/2}}{4} G_i T^i \label{eq:W2} \; .
\eee
We shall now consider the idea from Refs.~\cite{Acharya:2000ps, Beasley:2002db}
that each term in \eqn{W1} is `sourced' by the wrapped M5-brane
and M2-brane respectively. 
From \eqn{dstarF}---the equation of motion for the flux---it can be shown that
\be
\int_{X_7} \left( *K + \frac{1}{2} A \wedge G \right) = 0
\Rightarrow G_i A^i = - VK \label{eq:GiAi} \; ,
\ee
where $V$ is the volume of the compact space as defined in \eqn{defVol}.
Also, as $G$ is harmonic, the projector of its singlet must be
constant over $X_7$ and so
\be
G \cdot \Phi = \frac{1}{V} \int_{X_7} \sqrt{-g} G \cdot \Phi
\Rightarrow G_i \varphi^i = \frac{1}{24} V G \cdot \Phi \label{eq:Giphii} \; .
\ee
We can thus substitute \eqn{GiAi} and \eqn{Giphii} into \eqn{W2} 
to rewrite the superpotential as
\be
W = \frac{7^{3/2} V }{4} \left( \frac{1}{24} G \cdot \Phi - i K \right) \; .
\ee
Clearly, this expression contains a term proportional to the singlet
of $G$ as well as one proportional to $K$. As we saw in section
\ref{sec:brane}, the wrapped M5-brane acts as a source for the singlet of
$G$ and the M2-brane acts as a source for $K$. Therefore, we can
interpret the moduli superpotential for the 4-dimensional effective theory
associated with M-theory on $G_2$ manifolds as being sourced by
contributions from the wrapped M5-brane and the M2-brane.

The Kahler potential for this theory and its derivatives with respect
to the superfields is given by
\bee
\kahler & = & -3 \ln \left( 7V \right) \label{eq:kv}\\
\Rightarrow \kahler_i & = & \frac{\partial \kahler}{\partial T^i}
= \frac{-1}{2V} \int_{X_7} \pi_i \wedge \Phi 
= -2 \varphi_i \\
\Rightarrow \kahler_{i\bar{\jmath}} & = & \frac{1}{4V}
\int_{X_7} \pi_i \wedge * \pi_j \; .
\eee 
Using these expressions, we can write 
$G$ in terms of the dual set of 4-forms like
\be
G = \frac{1}{4V} G^i * \pi_i \; . \label{eq:gibar}
\ee

This is all the input from 11 dimensions that we need to write down and solve
the appropriate 4-dimensional Killing spinor equations. We will return to the links
between 4 and 11 dimensions later.

\subsection{4-dimensional Killing spinor equations}

We shall now set up the conditions for supersymmetric domain
wall solutions in the 4-dimensional supergravities derived from our
11-dimensional theory as above.
We start with a  warped metric ansatz
\be
ds_4^2 = e^{a(z)} \eta_{\mu \nu} dx^{\mu} dx^{\nu} + dz^2
\ee
and $z$-dependent scalar fields $T^i=T^i(z)$. For such a field
configuration the first order relations~\footnote{
There is one further equation that constrains the 4-dimensional
Killing spinor but it does not really impact on our calculation.}
derived from the Killing spinor equations are given by~ \cite{Eto:2003bn}
\bee
\partial_z a & = &
e^{-i \theta} e^{\frac{1}{2} \kahler} W
\label{eq:4Kfirst} \\
\partial_z T^i & = &
- e^{i \theta} e^{\frac{1}{2} \kahler} 
\kahler^{i\bar{\jmath}}\ D_{\bar{\jmath}} \overline{W}
\\
\partial_z \theta & = & - \mathrm{Im} \left[ (\partial_z T^i) \kahler_i \right]\; .
\label{eq:4Klast}
\eee
Here, $\theta$ parameterizes the 4-dimensional Killing spinor in a similar way
to the quantity $\theta$ we have used to parameterize the earlier 11-dimensional
spinor. When we consider the links between 4 and 11 dimensions later, we will
find that they are in fact the same at zero mode level in 11 dimensions.

Solutions to these equations are BPS states which preserve
half of the maximum number of supercharges possible in the $N=1$,
$D=4$ theory, and automatically satisfy its equations of motion.
Using the expressions for $T^i$, $W$ and $\kahler$ from 11 dimensions above
allows us to write the first-order relations as
\bee
\partial_z a & = & \frac{V^{-3/2}}{4} 
\left(G_i \varphi^i  \cos \theta + G_i A^i \sin \theta \right)
\label{eq:Adot}
\\
G_i A^i \cos \theta & = & G_i \varphi^i \sin \theta 
\\
\partial_z \varphi^i & = & \frac{V^{-3/2}}{4} \left( 
(2 G_j \varphi^j \varphi^i - G^i) \cos \theta
+ 2 G_j A^j \varphi^i \sin \theta \right) \label{eq:phidot}
\\
\partial_z A^i & = & \frac{V^{-3/2}}{4} \left( 
(2 G_j \varphi^j \varphi^i - G^i) \sin \theta 
- 2 G_j A^j \varphi^i \cos \theta \right) \label{eq:axdot} 
\\
\partial_z \theta & = & - \frac{V^{-3/2}}{2} G_i \varphi^i \sin \theta\; .
\label{eq:thetadot}
\eee
It is also worth noting that the relation
\be
\partial_z V = \frac{V^{-1/2}}{6} \left(
5 G_i \varphi^i \cos \theta + 7 G_i A^i \sin \theta \right) 
\label{eq:dzVol} \; .
\ee
can be derived from Eqs.~\eqn{Adot}--\eqn{thetadot}. This will be important later
when we find explicit solutions.

We now consider how to integrate these 4-dimensional equations in
purely 4-dimensional language, before uplifting them to compare with 
the 11-dimensional equations.

\subsection{Integrating the 4-dimensional equations}

We now present the most general solution to equations \eqn{4Kfirst}--\eqn{4Klast},
which are the conditions for a supersymmetric domain wall
to 4-dimensional SUGRA, given that the SUGRA descends from M-theory on a
$G_2$ manifold. The solution can be written, in terms of new moduli fields $f^i$, as
\begin{eqnarray}
 a & = & \frac{1}{2} \ln |\cot \theta | + C \\
 \varphi_i &=& \tan (\theta ) f_i \\
 A^i & = & \frac{-1}{4V_f \cot^{-7/3} \theta } G^i u  + b^i\; ,
\end{eqnarray}
where the new transverse coordinate $u$ is related to $z$ by
\begin{equation}
 \partial_u  =  ( V_f^{1/2} \cot^{1/2} \theta \mathrm{\, cosec\, } \theta ) \partial_z\; .
\end{equation}
Here, $C$ and $b^i$ are constants of integration. Recall, that once the
Kahler potential is explicitly given, the volume $V$ is known, via Eq.~\eqn{kv},
as a function of the moduli $\varphi^i$. By $V_f$ we mean this function but
with the fields $\varphi^i$ replaced by $f^i$. Since the volume is a homogeneous
function of degree $7/3$ in its arguments this means $V$ and $V_f$ are related
by
\begin{equation}
 V_f = V\cot^{7/3}\theta\; .
\end{equation}
The angle $\theta$ is fixed by
\begin{equation} 
 \cos \theta \cot^6 \theta  =  \left( \frac{V_f}{V_0} \right)^{3/2}\; ,
\end{equation}
where $V_0$ is another constant of integration. Finally, the new field $f_i$
are linear functions
\begin{equation}
 f_i  = \frac{1}{4} G_i w  + k_i\; ,
\end{equation}
in the transverse coordinate $w$ defined by
\begin{equation}
 \partial_w  = ( V_f^{3/2} \tan^{9/2} \theta \sec \theta ) \partial_z\; ,
\end{equation}
and $k_i$ are further constants of integration. This completes the most
general domain wall solution to 4-dimensional $N=1$ supergravity theories
from M-theory flux compactifications on $G_2$ spaces.

Note that the above solutions display some apparent singularities in the cases
$\sin\theta = 1$ or $\cos\theta =1$ which we have previously encountered when
matching to membrane and five-brane sources. However, these singularities are not
real but can be removed by introducing a new quantity $n$ defined by either
one of the two equivalent relations
\begin{equation}
 n =  \frac{3 \mathrm{ln} | \cot \theta |}{7 \mathrm{ln}
| \cos \theta | - 5 \mathrm{ln} | \sin \theta |}\; ,\qquad
\left(\frac{V}{V_0}\right)^n  = \left| \mathrm{cot}\, \theta \right| \; .
\end{equation}
In the above solution, we can in general eliminate $\theta$ in favour of $n$.

In this new form one can explicitly consider the cases $\sin\theta =1$ and
$\cos\theta =1$ without encountering any singularities. They lead to the
particularly simple solutions
\be
\begin{array}{|c|c|}
\hline
\cos \theta = 1 & \sin \theta = 1 \\
\hline
\partial_w = V^{9/10} \partial_z & \partial_u = V^{1/2} \partial_z \\
f_i = V^{3/5} \varphi_i = \frac{1}{4v} G_i w + k_i &
f_i = V^{3/7} \varphi_i = \mathrm{const.} \\
A^i = \mathrm{const.} &
A^i = - \frac{1}{4V} G^i u + b^i \\
a = a_0 + \mathrm{ln} V^{3/10} & a = a_0 + \mathrm{ln} V^{3/14}
\\ \hline
\end{array}
\ee
The results in this section represent the most general
supersymmetric domain wall solution to 4-dimensional supergravities
that arise from compactification of M-theory on $G_2$ manifolds with
flux. In particular, the solution with $\cos\theta =1$ in the above
table is the appropriate one to match to a wrapped 5-brane source
while the solution for $\sin\theta =1$ can be matched to a membrane
source. We note the linear dependence of the moduli fields on the
natural transverse coordinate $w$ or $u$, which will have the effect
of causing fields to diverge at large distances. In particular, the
solutions do not approach four-dimensional Minkowski space
asymptotically. This behaviour is typical for supergravity domain
walls supported by potentials without a minimum at any finite
field value~\cite{Cowdall:1996tw}. We will discuss curvature
singularities in section \ref{sec:4dsing} below.

\subsection{Comparison with 11 dimensions}

As a check of our results, we shall now test for compatibility between
the 4- and 11-dimensional relations derived from the 
Killing spinor equations.  Our strategy will
be to link the 4- and 11-dimensional quantities and then
uplift the 4-dimensional equations to 11 dimensions,
and see if they agree.

We firstly relate the quantities.
By comparing the 4-dimensional and 11-dimensional line elements 
$ds_4^2$ and $ds_{11}^2$ we see that
\bee
\partial_y & = & \pm \sqrt{V} \partial_z \label{eq:yz}
\\
\alpha & = & a - \frac{1}{2} \mathrm{ln} (V) \label{eq:alphaa} \; .
\eee
Using the result \eqn{indmet} of Appendix \ref{app:g2met} it is clear that
to first order
\be
\partial_y h_{AB} = \frac{1}{2} 
(P_{27} \partial_y \varphi)_{\left( A \right.}^{\ \ \ CD} 
\varphi_{\left. B \right) CD}
+ \frac{1}{63} ( \partial_y \varphi ) \cdot \varphi  \Omega_{AB} \; .
\ee
Making use of relations \eqn{Giphii} and \eqn{GiAi}, 
we simply substitute these results into the 4-dimensional bosonic equations
for appropriate sign choice in \eqn{yz} and get the following
\bee
\partial_y \alpha & = & \frac{1}{144} G \cdot \Phi \cos \theta
- \frac{1}{3} K \sin \theta \label{eq:4dcomp1st}
\\
\partial_y \mathrm{tr} (h) & = & - \frac{5}{72} G \cdot \Phi \cos \theta
+ \frac{7}{3} K \sin \theta
\\
\partial_y (P_{27} h)_{AB} & = & - \frac{1}{6} 
(P_{27} G)_{(A }^{\ \ \ CDE} \Phi_{B)CDE} \cos \theta
\\
\partial_y \theta & = & - \frac{1}{48} G \cdot \Phi \sin \theta
- K \cos \theta
\\
G \cdot \Phi & = & 24 K \cos \theta
\\
J \cdot \varphi & = & 21 K \cos \theta + \frac{5}{8} G \cdot \Phi \sin \theta
\label{eq:4dcomplast}
\eee
Although the equations linking components of the flux are superficially
different from the 11-dimensional ones, they are easily shown to be equivalent.

We note here that while the 4-dimensional equations did
not involve linearization in the flux while the 11-dimensional ones did.
Nevertheless, it turns out by comparing Eqs.~\eqn{4dcomp1st}--\eqn{4dcomplast} with
Eqs.~\eqn{zm1st}--\eqn{zmlast} that the two sets of equations are equivalent
at the zero mode level.

\subsection{Curvature singularities}

\label{sec:4dsing}

Another feature of the 4-dimensional equations that we should compare
with 11-dimensions is the variation of the volume as a function of the
transverse coordinate.  A zero of this volume at any particular point
in the transverse space implies, of course, a curvature singularity of
the internal space. However, it can be shown that such a vanishing
internal volume also leads to a four-dimensional curvature
singularity.

The $z$-variation of the volume is described by Eq.~\eqn{dzVol}
which uplifts to give the following
\be
\partial_y V =  \left[ - \frac{5}{144} G \cdot \Phi \cos \theta
+ \frac{7}{6} K \sin \theta \right] V \; . \label{eq:Vy}
\ee
This is the generalization to all orders in flux of the 11-dimensional result \eqn{dyVol} in
the zero-mode regime. When we consider the second derivative
of Eq.~\eqn{Vy}, using our 4-dimensional equations, we get
\be
\partial_y^2 V = \frac{-1}{12 V}
\left[
(G_i \varphi^i)^2 (2 \sin^2 \theta - 5)
+ \frac{1}{2} G_i G^i (2 \sin^2 \theta + 5)
\right] \; ,
\ee
which is quadratic in the flux. Writing this in 11-dimensional
language gives
\be
\partial_y^2 V = \frac{-1}{12 V} (2 \sin^2 \theta - 5)
\left[ \int_{X_7} \varphi \wedge G \right]^2
- \frac{1}{6} (2 \sin^2 \theta + 5) \int_{X_7} *G \wedge G \; .
\ee
This vanishes at linear order in flux which is consistent with our earlier findings
in Eq.~\eqn{dydyVol}. These implied a linear behaviour of the volume
and, therefore, its vanishing at some finite coordinate $y$.
However, the above result, good to all orders in flux, shows
that, in fact, the volume varies in a more complicated way.
In particular, the vanishing of the volume at some finite $y$
cannot be deduced generically at this stage.
To decide whether or not the volume vanishes one may study
specific examples of $G_2$ manifolds where $V$ is given as
an explicit function of the fields, as in Ref.~\cite{Lukas:2003dn}.

\section{Conclusions}

\label{sec:conc}

In this paper, we have studied M-theory compactifications on $G_2$ spaces
in the presence of flux, both from the viewpoint of the 11-dimensional
theory and the associated 4-dimensional supergravity theories.  
We have solved the 11-dimensional Killing spinor equations to linear
order in flux and obtained $G_2$ domain walls, consisting of a warped
product of a deformed $G_2$ space and a domain wall in four-dimensional
space-time. The zero mode parts of these solutions have also been
reproduced, to all orders in flux, by solving the Killing spinor equations
of the associated effective four-dimensional $N=1$ supergravity theories,
obtained by reducing M-theory on $G_2$ spaces with flux. From this
four-dimensional perspective, the solutions are domain walls
which couple to the flux superpotential and with moduli 
varying non-trivially along the transverse direction. This
transverse variation of the scalar fields can be seen as a path
in the moduli space of $G_2$ metrics or, in other words, a variation
of the internal $G_2$ space as one goes along the transverse direction.

We have also shown that these domain wall solutions can be sourced by
either a membrane in four-dimensional space-time or an M5-brane
wrapping a 3-cycle within the $G_2$ space. This leads to an
interpretation of our solutions as the simplest manifestation of an
M-theory ``topological defect'' membrane or wrapped 5-brane appearing
in ``our'' four-dimensional universe. We believe that studying such
defects, arising from wrapped branes, in the context of M-theory
cosmology is an interesting problem. 

Our four-dimensional domain walls diverge away from the wall and, in particular,
do not approach Minkowski space. This is a common feature of supergravity
domain walls~\cite{Cowdall:1996tw}. It would be interesting to see how these
solutions are modified by the inclusion of a non-perturbative
superpotential in four dimensions and whether this can remove the
divergences. We hope to return to this question in a future publication.

\section*{Acknowledgments} T.~H.~is supported by a PPARC studentship.
A.~L.~is supported by a PPARC Advanced Fellowship.

\vspace{1cm}

\appendix

\section{Spinor conventions}

\label{app:spin}

Throughout this paper we have made use of certain properties
of the Dirac matrices in various dimensions. We collect these
here for reference.

\subsection{Three dimensions}

In our conventions, the 3-dimensional Minkowski Dirac 
matrices are given by
\bee
\rho^{0} & = & -i \sigma^{2} \nonumber \\
\rho^{1} & = & \sigma^{1} \nonumber \\
\rho^{2} & = & - \sigma^{3} \; .
\eee
These obey the following
\bee
{\rho^{0} \rho^{1} \rho^{2}} & = & - \unitop_2
\\
( \rho^{\mu})^* & = & \rho^{\mu}
\\
\left\{ \rho^{\mu},\rho^{\nu}  \right\} & = & 2 \eta^{\mu \nu}
\eee
where $\mu, \nu = 0,1,2$.

\subsection{Four dimensions}

The 4-dimensional Dirac matrices are constructed from the above by
\bee
\gamma^{\mu} & = & \rho^{\mu} \otimes \sigma^{1} \nonumber \\
\gamma^{y} & = & \unitop_2 \otimes \sigma^{2} \; .
\eee
We can define a 4D chirality operator
$\gamma = i \gamma^0\gamma^1\gamma^2\gamma^y$ so that
\be
\left\{ \gamma, \gamma^y \right\} = 
\left\{ \gamma, \gamma^{\mu} \right\} = 
\left\{ \gamma^{\mu}, \gamma^y \right\} = 0
\ee
\be
\left\{ \gamma^{\mu},\gamma^{\nu}  \right\} = 2 \eta^{\mu \nu}
\ee
\be
(\gamma)^2 = ( \gamma^y )^2 = 1 \; .
\ee
These matrices further obey the following
\bee
\hat{\varepsilon}_{\mu \nu \rho} \gamma^{\mu} \gamma^{\nu} \gamma^{\rho}
& = & 6i \gamma^y \gamma
\\
\hat{\varepsilon}^{\mu}_{\ \nu \rho} \gamma^{\nu} \gamma^{\rho} & = &
2i \gamma^y \gamma \gamma^{\mu}
\eee
\be
\begin{array}{ccc}
(\gamma^{\mu})^* = \gamma^{\mu} & 
(\gamma)^* = \gamma&
(\gamma^{y})^* = - \gamma^{y}
\end{array} \; ,
\ee
where $\hat{\varepsilon}_{012} := 1$.

\subsection{Seven dimensions}

In 7-dimensional  Euclidean space it is possible to define a set of
Dirac matrices $ \{ \Upsilon^A \} _{A = 4 \ldots 10} $ that
are purely imaginary so that
\bee
(\Upsilon^A)^* & = & - \Upsilon^A
\\
\left\{ \Upsilon^A, \Upsilon^B \right\} & = & 2 \delta^{AB} \; .
\eee

\subsection{11 dimensions}

Finally, we define the set of 11-dimensional Dirac matrices
\be
\left\{ \Gamma^I \right\} _{I = 0 \ldots 10}= 
\left\{ \Gamma^{\mu}, \Gamma^y, \Gamma^A \right\}
_{\mu = 0,1,2, A = 4 \ldots 10}
\ee
by the relations
\bee
\Gamma^{\mu} & = & \gamma^{\mu} \otimes \unitop_8 \nonumber \\
\Gamma^{y} & = & \gamma^{y} \otimes \unitop_8 \nonumber \\
\Gamma^{A} & = & \gamma \otimes \Upsilon^A \; .
\eee
Throughout this paper we use the standard notation
\be
\Gamma^{I_1 \ldots I_p}  = 
\Gamma^{\left[ I_1 \ldots \right.} \Gamma^{\left. I_p \right]} \; ,
\ee
for anti-symmetrized products of Dirac matrices. 

\subsection{Majorana Conjugation}

Since 11-dimensional supergravity is parameterized by Majorana spinors,
it is useful to be clear about our conventions for charge conjugation
and, in view of our compactification, how this operation decomposes
under a product spinor Ansatz.

For a general spinor $\psi$, we define its Majorana conjugate
in terms of the matrix $B$
\be
\psi^c = B^{-1} \psi^* \; .
\ee
Imposing that this operation should commute with Lorentz transformations
and square to unity gives the conditions
\be
\begin{array}{cc}
B\Gamma^I B^{-1} = \pm (\Gamma^I)^*  & B^* B = \unitop \; . \label{eq:maj}
\end{array}
\ee
Let us decompose $B$ as $B^{-1} = B_4^{-1} \otimes B_7^{-1}$
into 4- and a 7-dimensional conjugation matrices $B_4$ and $B_7$.
They must satisfy the relations
\be
\begin{array}{cc}
B_{4}\gamma^{m} B_{4}^{-1} = \pm (\gamma^{m})^*
& B_{7} \Upsilon^A B_{7}^{-1} = \mp (\Upsilon^A)^*
\end{array}
\ee
where $m = 0,1,2,y$, in order to reproduce Eq.~\eqn{maj}.
In our conventions, these matrices can be represented as
\be
\begin{array}{cc}
B_{4}^{-1} = \gamma^y \gamma
& B_{7}^{-1} = \unitop_8 \; .
\end{array}
\ee

\section{Properties of $G_2$}

\label{app:g2}

We shall now state some results for manifolds of $G_2$ holonomy,
in part following Ref.~\cite{joyceb,Bilal:2001an}.

\subsection{$G_2$ Lie Algebra} \label{app:g2lie}

In this section we define $G_2$ as a subgroup of ${\rm SO}(7)$ by
defining the $G_2$ 3-form $\varphi$. We also decompose the
Lie algebra of ${\rm SO}(7)$ into a part in $G_2$ and a perpendicular
part.

${\rm SO}(7)$ and its Lie algebra are given by
\bee
{\rm SO}(7) & := & \left\{ O \in  Gl(7, \Reals) \left| 
O = O^T\ \& \ \mathrm{det}(O) = 1 \right. \right\} \\
\Rightarrow \mathscr{L} ({\rm SO}(7)) & = & \left\{ T
\left| T^T = -T \right. \right\} \nonumber \\
 & = & \mathrm{Span}\left\{ S_{AB} \right\} \; .
\eee
where the basis of 21 generator matrices is
\be
(S_{AB})^C_{\ D} := \delta^C_A\delta_{BD} - \delta_{AD}\delta^C_B \; .
\ee
$G_2$ is then defined as the subgroup of ${\rm SO}(7)$ whose action
preserves the 3-form $\varphi$
\be
\varphi = d\mathbf{x}^{123} + d\mathbf{x}^{145}
 + d\mathbf{x}^{167} + d\mathbf{x}^{246} 
- d\mathbf{x}^{257} - d\mathbf{x}^{347}
- d\mathbf{x}^{356}
\ee
where $d\mathbf{x}^{A_1 \ldots A_p} 
:= dx^{A_1} \wedge \ldots \wedge  dx^{A_p}$.
$G_2$ also preserves the Hodge dual of this 3-form,
$\Phi := * \varphi $. We may thus write
\bee
G_2 & := & \left\{ P \in {\rm SO}(7) \left| P^{A'}_{\ A}P^{B'}_{\ B}P^{C'}_{\ C}  
\varphi_{A' B' C'} = \varphi_{ABC} \right. \right\} \\
\Rightarrow \mathscr{L} (G_2) & = & \left\{ T \in \mathscr{L} ({\rm SO}(7))
\left| T^{A'}_{\ A} \varphi_{A' BC} + T^{B'}_{\ B} \varphi_{A B' C}
T^{C'}_{\ C} \varphi_{ABC'} = 0 \right. \right\} \; . \nonumber \\
\eee
We can then split the ${\rm SO}(7)$ generators into a group of 7 not in
$\mathscr{L}(G_2)$ and a group of 14 in $\mathscr{L}(G_2)$, which
are given by
\bee
F_A & = & \frac{1}{2} \varphi_{A}^{\ \ BC} S_{BC} \\
R_{AB} & = & \frac{2}{3} S_{AB} - \frac{1}{6} \Phi_{AB}^{\ \ \ \ CD} S_{CD}
\eee
respectively. This represents the branching rule
\be
\irrep{21}_{{\rm SO}(7)} \rightarrow (\irrep{14} + \irrep{7})_{G_2} \; .
\label{eq:adjdecomp}
\ee

\subsection{Projector conventions} \label{app:project}

In solving equations containing $G_2$ spinors, vectors, and
tensors, it is usually easiest to project out irreducible
representations of the relevant indices. Here we summarize
our conventions for the projectors that we use. \\
For a 2-form, $\xi$, decomposing as $({\bf 7}\times{\bf 7})_{\rm anti-symm.}\rightarrow
{\bf 7}+{\bf 14}$, we have
\bee
(P_7 \xi )_{AB} & = & \frac{1}{6} (\xi_{CD} \varphi^{CDE} \varphi_{EAB} )
\\
P_{14} & = & \unitop - P_7 \; .
\eee
For a 3-form, $\zeta$, decomposing as $({\bf 7}\times{\bf 7}
\times{\bf 7})_{\rm anti-symm.}\rightarrow{\bf 1}+{\bf 7}+{\bf 27}$ we have
\bee
(P_1 \zeta)_{ABC} & = & \frac{1}{42} (\varphi \cdot \zeta )\varphi_{ABC}
\\
(P_7 \zeta)_{ABC} & = & \frac{1}{24} \Phi^{DEFG} \zeta_{EFG} \Phi_{DABC}
\\
P_{27} & = & \unitop - P_1 - P_7 \; .
\eee
For a 4-form, $\chi$, decomposing as $({\bf 7}\times{\bf 7}
\times{\bf 7}\times{\bf 7})_{\rm anti-symm.}\rightarrow{\bf 1}+{\bf 7}+{\bf 27}$ we have
\bee
(P_1 \chi )_{ABCD} & = & \frac{1}{168} = (\Phi \cdot \chi ) \Phi_{ABCD}
\\
(P_7 \chi )_{ABCD} & = & \frac{1}{42} \varphi^{EFG} 
\chi_{EFG \left[ A \right. } \varphi_{\left. BCD \right] }
\\
P_{27} & = & \unitop - P_1 - P_7 \; .
\eee
For a symmetric rank 2 tensor, $S$, decomposing as $({\bf 7}\times{\bf 7})_{\rm symmetric}
\rightarrow{\bf 1}+{\bf 27}$, we have
\bee
(P_1 S)_{AB} & = & \frac{1}{7} S^C_{\ C} \Omega_{AB}
\\
P_{27} & = & \unitop - P_1 \; .
\eee
For an ${\rm SO}(7)$ spinor, decomposing under $G_2$ as ${\bf 8}_{{\rm SO}(7)}\rightarrow
({\bf 1}+{\bf 7})_{G_2}$ we have 
\bee
P_1 & = & \frac{1}{8} \left( \unitop - \frac{1}{4!} \Phi_{ABCD} \Upsilon^{ABCD}
\right)
\\
P_7 & = & \unitop - P_1 \; .
\eee

\subsection{$G_2$ spinors} \label{app:g2spin}

In this section, we construct a basis set for general spinors on a $G_2$
manifold using some of the results from earlier appendices. We also
consider the action of our Dirac matrices on these spinors, which
is crucial in solving the Killing spinor equations.

The 7-dimensional Dirac matrices can be used to provide a representation
of $\mathscr{L}({\rm SO}(7))$ via
\be
\Sigma_{AB} := \frac{1}{2} \Upsilon_{AB}
\ee
which obey the same commutation relations as the $S_{AB}$ above.
We can use this isomorphism to split the $\Sigma_{AB}$ as for Appendix
\ref{app:g2lie}
\bee
f_A & = & \frac{1}{2} \varphi_{A}^{\ \ BC} \Sigma_{BC} \\
\rho_{AB} & = & \frac{2}{3} \Sigma_{AB} 
- \frac{1}{6} \Phi_{AB}^{\ \ \ \ CD} \Sigma_{CD} \; .
\eee
We are then able to define a `zeroth order' spinor $\chi_0$ by
\be
\begin{array}{ccc}
P_1 \chi_0 = \chi_0 & P_7 \chi_0 = 0 & \overline{\chi_0 } \chi_0 = 1
\end{array}
\ee
which is covariantly conserved on the $G_2$ manifold,
i.e. $\rho_{AB} \chi_0 = 0$.
We further define a set of seven other spinors $\{ \chi_A \}$ obeying
\be
\begin{array}{rclrcl}
\chi_A & := & \frac{2}{3} f_A \chi_0 & P_7 \chi_A & = & \chi_A \\
\overline{\chi_0} \chi_A & = & \overline{\chi_A} \chi_0 = 0 &
\overline{\chi_A} \chi_B & = & \delta_{AB}
\end{array}
\ee
which represents the branching rule
\be
\mathbf{8}_{{\rm SO}(7)} \rightarrow \left( \mathbf{7 + 1} \right)_{G_2} \; .
\label{eq:spindecomp}
\ee
We therefore have $\{ \chi_0, \chi_A \}$ as
a basis for spinors on a $G_2$ manifold.
The action of our Dirac matrices on these spinors is then given by
\bee
\Upsilon^A \chi_0 & = & i \Omega^{AB} \chi_{B} \\
\Upsilon^A \chi_B & = & -i \delta^A_B \chi_0
+i \varphi_{B}^{\ \ AC} \chi_C \\
\Upsilon^{AB} \chi_0 & = & \varphi^{ABC} \chi_C \\
\Upsilon^{AB} \chi_C & = & - \varphi^{AB}_{\ \ \ C} \chi_0
+ ( \Phi_C^{\ \ ABD} + \delta^B_C \Omega^{AD} 
- \delta^A_C \Omega^{BD} ) \chi_D \\
\Upsilon^{ABC} \chi_0 & = & -i \varphi^{ABC} \chi_0
+i \Phi^{ABCD} \chi_D \\
\Upsilon^{ABCD} \chi_0 & = & - \Phi^{ABCD} \chi_0 -
4 \varphi^{\left[ ABC \right. }\Omega^{\left. D \right] E } \chi_E\\
\Upsilon^{ABCDE} \chi_0 & = & -i \left(
\Phi^{\left[ ABCD \right.} \Omega^{\left. E \right] F} +
4 \varphi^{\left[ ABC \right.} \varphi^{\left. DE \right] F} \chi_F \right)
\eee 
where $\varphi$ is the $G_2$ 3-form, $\Phi$ is its dual, $\Omega$ is the
metric on the $G_2$ space and the $\Upsilon$ are the 7-dimensional Dirac matrices.

\subsection{$G_2$ induced metric} \label{app:g2met}

The $G_2$ form $\varphi$ naturally induces a metric on a $G_2$ manifold

\bee
\Omega_{AB} & = & \left( \mathrm{det} (s) \right)^{-1/9} s_{AB} 
\mathrm{\ for} \nonumber \\
s_{AB} & = & \frac{1}{144} \varphi_{ACD}\varphi_{BEF}\varphi_{GHI}
\hat{\varepsilon}^{CDEFGHI}
\eee
where the entries of the completely antisymmetric
tensor $\hat{\varepsilon}$ are either 1 or 0.
By writing $\varphi = \varphi^i \pi_i$ it is then possible to
show that
\be
\partial g_{AB} = \left[ \frac{1}{2} 
\varphi_{\left( A \right.}^{\ \ \ CD} (\pi_i)_{\left. B \right) CD}
- \frac{1}{18} ( \varphi \cdot \pi_i ) g_{AB}
\right] \partial \varphi^i \label{eq:indmet}
\ee
using some of the identities below.

\subsection{$G_2$ identities}

Throughout this paper, the following $G_2$ identities have been used:
\bee
\varphi \cdot \varphi & = & 42 \\
\varphi^{ACD} \varphi_{BCD} & = & 6 \delta^A_B \\
\varphi_{ABE} \varphi_{CD}^{\ \ \ \ E} & = &
\Phi_{ABCD} + \Omega_{AC}\Omega_{BD} - \Omega_{AD}\Omega_{BC} \\
\Phi \cdot \Phi & = & 168 \\
\Phi^{ACDE} \Phi_{BCDE} & = & 24 \delta^A_B \\
\Phi_{ABEF} \Phi_{CD}^{\ \ \ \ EF} & = & 2\Phi_{ABCD}
+ 4 (\Omega_{AC}\Omega_{BD} - \Omega_{AD}\Omega_{BC}) \\
\varphi^{BCD} \Phi_{ABCD} & = & 0 \\
\varphi_{A}^{\ \ DE} \Phi_{BCDE} & = & 4 \varphi_{ABC} \\
\varphi_{AB}^{\ \ \ \ F} \Phi_{FCDE} & = & 
\Omega_{AC} \varphi_{BDE} + \Omega_{AD} \varphi_{BEC} 
+\Omega_{AE} \varphi_{BCD} - \nonumber \\
& & \Omega_{BC} \varphi_{ADE} 
- \Omega_{BD} \varphi_{AEC} - \Omega_{BE} \varphi_{ACD} \\
\varphi_{\left[ ABC \right.} \Phi_{\left. DEFG \right] } & = & 
\frac{1}{5} \varepsilon_{ABCDEFG}
\eee
\be
3 \varphi_{(A}^{\ \ \left[ CD \right. } 
\varphi^{\left. EF \right] }_{\ \ \ \ B) }
+ 4 \Phi^{\left[ CDE \right. }_{\ \ \ \ \ \ (A } 
\delta^{\left. F \right] }_{B)}
- \Phi^{CDEF} \Omega_{AB} = 0
\ee
\be
\epsilon^{CDEFGHI} \varphi_{ACD} \varphi_{BEF} = 
24 \varphi^{\left[ GH \right. }_{\ \ \ \ (A } \delta^{\left. I \right]}_{B)} \; ,
\ee
where $\varphi$ is the $G_2$ 3-form, $\Phi$ is its dual and $\Omega$ is the
metric on the $G_2$ space.
For a 4-form $\alpha$ such that $P_7 \alpha = 0$, we have the further
results that
\bee
\alpha \cdot \Phi & = & 4 (* \alpha) \cdot \varphi
\\
{(P_{27} \alpha)_{(A}}^{CDE} \Phi_{B)CDE} 
& = & -3 {(P_{27} * \alpha)_{(A}}^{CD} \varphi_{B)CD}\; .
\eee

\end{document}